\documentclass[11pt,fleqn]{article}
\usepackage{amsmath,amssymb,amsthm,epsfig,graphics}

\newtheorem*{lemma*}{Lemma}

\newtheorem*{conjecture*}{Conjecture}

{\theoremstyle{definition}

\newtheorem*{note*}{Note}
}

\allowdisplaybreaks

\begin{document}

\begin{center}
\LARGE \bf
Qusicryslals in Vernam cipher
\end{center}

\begin{center} \bf
Maryna NESTERENKO~$^\dag$ and Severin PO\v STA~$^{\S}$
\end{center}

\noindent $^\dag$~Institute of Mathematics of NAS of Ukraine, 3
Tereshchenkivs'ka Str., 01004, Kyiv, Ukraine\\
National Technical University of Ukraine "Igor Sikorsky Kyiv Polytechnic Institute"\\
37 Beresteyskyi av., 03056, Kyiv, Ukraine
$\phantom{^\dag}$~E-mail: maryna@imath.kiev.ua, severin.posta@fjfi.cvut.cz

\noindent $^\S$~Faculty of Nuclear Sciences and Physical Engineering, Czech Technical University in Prague, 13 Trojanova str., 12000, Prague, Czech Republic


\begin{abstract}
\noindent
We propose a modification of the classical Vernam cipher based on
properties of one-dimensional quasicrystals.
The method uses the sequence of quasicrystal minimal distances as an one-time pad.
The main advantages are strict aperiodicity of such sets
and rather short key that uniquely defines quasicrystal.
\end{abstract}

\section*{Introduction}
The Vernam cipher was first described by Frank Miller in 1882~\cite{Bellovin2011}, but named after Gilbert Vernam, who in 1917 invented and in 1919 patented~\cite{Vernam1919} a system for automatically encrypting teletype messages. Much later in 1945 it was proven by Shannon Claude~\cite{Claude1949} that the Vernam cipher is is cryptographically absolutely secure.
That is, intercepting an encoded data without a key does not provide any information about the initial data. From the point of view of cryptography, it is impossible to come up with a system more secure than the Vernam cipher.
As of now, the Vernam cipher is used quite rarely due to the significant size of the key, the length of which must match the length of the message, that is, the use of such ciphers requires enormous costs for the production, storage and transition of keys. Nevertheless, completely secure ciphers still found practical application for protecting particularly important communication lines with a relatively small amount of information and for application in post-quantum cryptography since there is a quantum analogue of a one-time pad that can be used to exchange quantum states over a one-way quantum channel with complete secrecy, which is sometimes used in quantum computing. Moreover, cryptographic algorithms that could be broken with a powerful enough quantum computer are no longer safe.
Namely, encryption methods based on mathematical quantities, the finding of which would take an incredibly long time with currently available computers, can be solved much faster by a future quantum computer. This means that current security algorithms can become vulnerable.
One-time pads, however, would remain secure, as perfect secrecy does not depend on the attacker's computational resources.

In this work, we propose a modification of the Vernam cipher, which consists in replacing the two-bit key with a two-bit sequence that corresponds to the minimum distances between the points of a one-dimensional quasi-crystal. This allows to overcome several main disadvantages of the classical cipher.

The paper is arranged as follows. First, we recall in Section~\ref{sec_Vernam} the essential part of the Vernam cipher and discuss its main points and disadvantages.
In Section~\ref{sec_quasicrystal} we present mathematical and computational ingredients that concern one-dimensional quasicrystals.
This is followed by the modification of Vernam cipher and description of its key.

\pagebreak
\section{Original Vernam cipher}\label{sec_Vernam}
Vernam's original cryptosystem was proposed for encrypting teletype messages, that is for texts.
The plaintext was first represented in binary form, after that each of the characters $1$ and $0$ was encoded using the exclusive or operation "XOR":
In this work, we will focus only on the core of the Vernam cipher and will not pay attention to the encryption of the text itself (assigning a binary number to each character and so on), since any digital data can be presented in binary form, or encryption can be applied to black and white pixel images, etc.

Note, that XOR operation in mathematical logics is usually called Zhegalkin addition or summation modulo 2 and is denoted by "$\oplus$", below we give the truth table for this operation and its negation --- the equivalence operation "$\sim$".
\begin{center}
\begin{tabular}{|c|c|c|c|}
\hline
$x$ & $y$ & $x\oplus y$ &$x\sim y$\\
\hline
0& 0 & 0 & 1\\
\hline
0& 1 & 1 & 0\\
\hline
1& 0 & 1 & 0\\
\hline
1& 1 & 0 & 1\\
\hline
\end{tabular}
\end{center}
Let us denote the initial data by the symbols $d_1$ and $d_2$, if we encode both this symbols using the same key $k$, then the attacker, having received two results of encryption $r_1$ and $r_2$, will immediately get additional information, namely $(d_1\oplus k) \oplus (d_2 \oplus k)=(d_1\oplus d_2) \oplus ( k\oplus k)=r_1\oplus r_2$ and
$d_1\sim d_2=\overline{r_1\oplus r_2}$.
This information significantly contributes to decryption, especially in the case of a text that is composed from the words of a certain language. These considerations prove that the key should not be reused.
Let's summarize the main requirements and shortcomings of the classic Vernam cipher:
\begin{itemize}
  \item the key should be random and can not be generated via some fixed algorithm, that expands one or more small values into a longer key;
  \item key reuse compromises security;
  \item key storage and transmission problem (if the key is encrypted with a non-information theoretically secure algorithm for delivery, the security of the cryptosystem is only as secure as the insecure delivery mechanism);
  \item practicality and efficiency (the pad must be destroyed immediately after use, Vernam ciphers are often impractical and used only in scenarios where complete confidentiality is most important);
  \item key length is equal to message length.
\end{itemize}
Our main goal is to offer such one-time pads that are uniquely fixed with a small number of random integers,
thereby reducing the amount of information that needs to be transmitted or stored in complete secrecy, and that can even be remembered by human without recording.

\section{Construction and properties of one-dimensional quasicrystals}\label{sec_quasicrystal}
It is possible to generate $d$-dimensional quasicrystals ($d=1,2,\dots$) (strictly aperiodic discrete sets) by three main approaches: tiling, deflation and cut-and-project method.
In this work, we will apply the cut-and-project method and focus on the construction of quasicrystals of the smallest dimensions, namely, one-dimensional ones.
Theoretical background of these model sets can be found in~\cite{Gazeau-Masakova-Pelantova2006,
Masakova-Patera-Pelantova1998a, Masakova-Patera-Pelantova1998c, MPP, Moody-Patera1998}.

\begin{figure}[ht]
\centerline{
\includegraphics[scale=0.6]{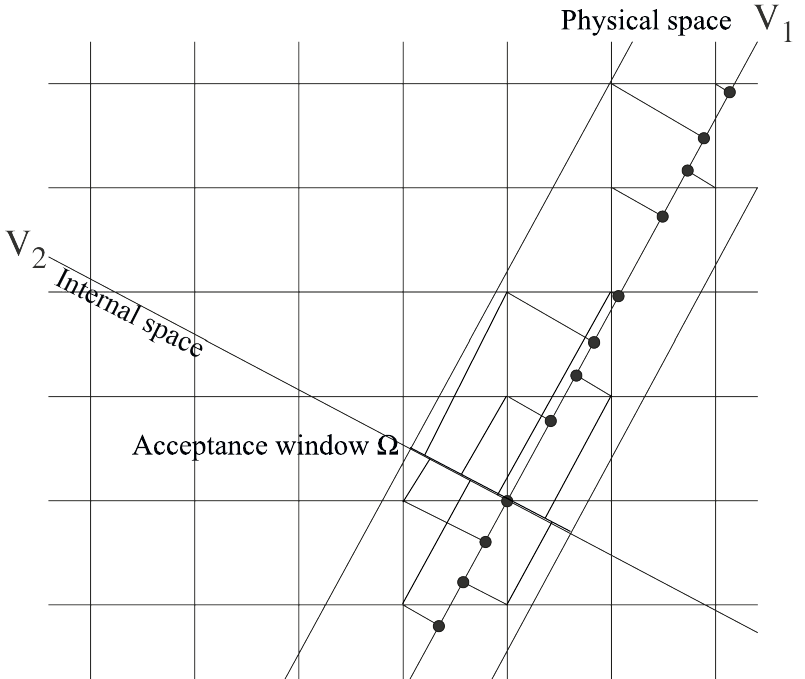}}
\caption{Construction of one-dimensional quasicrystal from $\mathbb{Z}_2$ lattice}\label{fig_cut-project}
\end{figure}

Our aim is to start with two-dimensional lattice and to construct a discrete aperiodic one-dimensional set.
We start with a choice of two straight lines $V_1\colon y=\tau x$ and $V_2\colon y=\tau' x$
on a~square lattice ${\mathbb Z}_2$ (see Fig.~\ref{fig_cut-project}),
where $\tau$ and $\tau'$ are two irrational numbers (it is the irrationality of the coefficients that guarantees the aperiodicity of discrete sets).

There are infinitely many choices of irrational pairs $\tau$ and $\tau'$,
Namely, one may take as $\tau$ and $\tau'$
the solutions of the two infinite series of the quadratic equations $x^2=mx+1,\ m=1,2,\ldots$
or $x^2=mx-1,\ m=3,4,\ldots$. The choice of equation must be random.
Note, that lines $V_1$ and $V_2$ are orthogonal in this cases.

One of the lines, say $V_1$, plays the role of the physical space onto which a `cut' of the  lattice ${\mathbb Z}_2$ is projected to obtain quasicrystal.
In the other direction $V_2$, we choose a finite interval $\Omega$.
Point of the lattice ${\mathbb Z}_2$ becomes a~point of the quasicrystal, after projection on $V_1$, provided its projection on $V_2$ falls within $\Omega$, i.e. it is within the `cut'.

We denote by ${\mathbb Z}[\tau]$ the extension of integer numbers, i.e. all the real numbers of the form $(a+b\tau)$, with integers $a$ and $b$.

The map $(\cdot)'$ between a quasicrystal point $(a+b\tau)$ and the point $(a+b\tau')$ of $\Omega$ is given by
\begin{gather*}
\begin{tabular}{rcl}
$(\cdot)'\ \colon\quad {\mathbb Z}[\tau]$ &$\longrightarrow$&${\mathbb Z}[\tau']$,\\
$(a+b\tau)$&$\mapsto$& $(a+b\tau')$,\qquad $a,b\in{\mathbb Z}$.
\end{tabular}
\end{gather*}

Let $\Omega$ be a finite interval called the \textit{acceptance window} (see Fig.~\ref{fig_cut-project}) and $\tau$, $\tau'$
be irrational numbers, then
\begin{gather*}
\Lambda_{\tau}(\Omega)=\{a+b\tau\mid a,b\in{\mathbb Z},\ a+b\tau'\in\Omega\},
\end{gather*}
is called the \textit{cut and project set} or one-dimensional \textit{quasicrystal} or \textit{model set}.

Discrete point set $\Lambda_{\tau}(\Omega)$ defined here is an infinite, uniformly dense, and uniformly discrete point set, while its window $\Omega$ is a finite interval densely covered by the star map of the points of $\Lambda_{\tau}(\Omega)$.

From the definition of $\Lambda_{\tau}(\Omega)$ a number of properties of $\Lambda_{\tau}(\Omega)$
can be shown.
In particular, the aperiodicity, discontinuity of the star map and the existence of only two or three different distances
between adjacent points of $\Lambda_{\tau}(\Omega)$.
For our applications, we will need the following statements.

Let $\Omega$ be a semi-closed interval. For every $\Lambda_{\tau}(\Omega)$ there
exist positive numbers $\Delta_1,\Delta_2\in {\mathbb Z}[\tau]$,
such that the distances between adjacent points in $\Lambda_{\tau}(\Omega)$
take values in $\{\Delta_1,\Delta_2,\Delta_1+\Delta_2\}$.
The distances depend only on $\tau,\tau'$ and length of the acceptance window, namely
\begin{gather*}
\{x_{n+1}-x_n|n\in{\mathbb Z}\}=
\left\{
\begin{array}{l}
\{\Delta_1,\ \Delta_2,\ \Delta_1+\Delta_2\}, \\
\text{when}\quad \Delta_1'-\Delta_2'>|\Omega|
\\[1 ex]
\{\Delta_1,\ \Delta_2\},\\
\text{when}\quad \Delta_1'-\Delta_2'=|\Omega|.
\end{array}
\right.
\end{gather*}

One has a general scaling property of quasicrystals:
For every irrational numbers $\tau,\tau'$, $\tau\ne\tau'$ and a bounded interval $\Omega$,
there exist $\tilde\tau'\in(-1,0)$, $\tilde\tau>0$, $s\in{\mathbb R}$ and
$\tilde\Omega$ (satisfying $\max\{1+\tilde\tau',-\tilde\tau'\}<|\tilde\Omega|\le 1$),
such that
$
\Lambda_{\tau}(\Omega)=s\Lambda_{\tilde\tau}(\tilde\Omega).
$

We can scale up any model set in such a way that the distances between adjacent points in $\Lambda_{\tilde\tau}(\tilde\Omega)$ are equal to
$\tilde\tau$, $1+\tilde\tau$, $1+2\tilde\tau$, when $|\tilde\Omega|\ne 1$ and to
$\tilde\tau$, $1+\tilde\tau$, when $|\tilde\Omega|=1$.

Finally we underline that the 1-dimensional quasicrystal $\Lambda_{\tau}(\Omega)$ is an infinite discrete
aperiodic set of points uniquely determined by the random choice of $\tau$, $\tau'$ and  of
acceptance window $\Omega=[c,c+d)$, where $c$ and $d$ are random  real numbers.
Without loss of generality, we can consider
$\tau'\in(-1,0)$, $\tau>0$, $c=0$ and
$d\in(\max\{1+\tau',-\tau'\},1]$.
Therefore, using the properties of one-dimensional quasi-crystals, we can avoid building projections on physical and internal space, and instead check whether the conjugate number belongs to the acceptance window.

It is important to note that the constructed quasicrystals are infinite-dimensional sets, and for encryption we always need finite one-time notebooks,
even if they are quite long.
Thus, knowing the length of the binary sequence to be encrypted, we have an infinite set of keys in each fixed quasi-crystal.
The selection of the key can be made randomly by fixing the seed point with two integers.
Let us consider two specific examples of one-dimensional quasicrystals.
\subsection{Example 1}
Take $\tau=\frac{1+\sqrt{5}}{2}\approx 1.62>0$, $\tau'=\frac{1-\sqrt{5}}{2}\approx -0.62\in(-1,0)$, the
two roots of the quadratic equation $x^2=x+1$. Then $\tau+\tau'=1$,
$\tau\tau'=-1$. Choosing $\Omega=[0,d)$ and $d=\frac{\tau}{2}$ we obtain that
the distances between adjacent points of the quasicrystal $\Lambda_{\tau}([0,d))$ are
$\Delta_1=\tau$, $\Delta_2=1+\tau$ and $\Delta_3=1+2\tau$.

As the seed point of the quasicrystal $\Lambda_{\tau}([0,d))$,
we can choose any point $x=a+b\tau$, such that $x'=a+b\tau'\in [0,d)$.
For our example we put $x=0$.

Now we have the information for fast generation of quasicrystal points.
One can move right or left from the seed point by adding or subtracting
one-by-one of the distances $\Delta_1$, $\Delta_2$ or $\Delta_3$.

Suppose we have already established, that $x$ is a point of $\Lambda_{\tau}([0,d))$, i.e. $x'\in [0,d)$.
The point adjacent to $x$ from the right is one of the three
$x+\tau,\quad x+1+\tau\quad \text{or}\quad x+1+2\tau$.

In order to decide which one is the case, we verify one-by-one the inclusions
$x+\tau'\in [0,d),\quad  x+1+\tau'\in [0,d)\quad \text{or}\quad x+1+2\tau'\in [0,d).$
The first confirmed inclusion determines the new quasicrystal point.

Depending on the order of adding the possible distances, the following two subsets of the quasicrystals
$\Lambda_{\tau}([0,d))$ were obtained.

First quasicrystal has the form
\begin{gather*}
\{\ldots, -1\!-\!2\tau,\ -\tau,\ 0,\ 1\!+\!\tau,\ 2\!+\!2\tau,\ 2\!+\!3\tau,\ 3\!+\!4\tau, \ldots\}
\end{gather*}

First quasicrystal rewritten in terms of minimal distances between quasicrystal points

\begin{gather*}
\{\ldots,\ \Delta_2,\ \Delta_1,\ \Delta_2,\ \Delta_2,\ \Delta_1,\ \Delta_2,\ \ldots\}.
\end{gather*}

First quasicrystal rewritten in terms of numbers for encoding
\begin{gather*}
\{\ldots,\ 1,\ 0,\ 1,\ 1,\ 0,\ 1,\ \ldots\}.
\end{gather*}

Second quasicrystal
$\{\ldots,-\tau,\ 0,\ 1+\tau,\ 2+3\tau,\ 3+4\tau,\ldots\}\leftrightarrow$
\begin{gather*}
\leftrightarrow\{\ldots,\Delta_1,\ \Delta_2,\ \Delta_3,\ \Delta_2,\ldots\}\leftrightarrow
\{\ldots,0,\ 1,\ 2,\ 1,\ldots\}.
\end{gather*}

Note, that the second quasicrystal can be applied for generalized Vernam cipher which corresponds to the summation of integers modulo three in the ring $\mathbb{Z}_3$
The two above quasicrystals differ by the sequence in which the three distances~$\Delta_1$, $\Delta_2$ and $\Delta_3$ were tried in the construction.

\subsection{Example 2}
This quasicrystal $\Lambda_{\tau}(\Omega)$ is built using the irrationalities $\tau=1+\sqrt{2}$, $\tau'=1-\sqrt{2}$,
roots of the quadratic equation $x^2=2x+1$.
Now $\tau'\approx-0.42\in(-1,0)$ and $\tau\approx 2.42>0$.
Putting $\Omega=[0,1)$,
there are only two distances $\Delta_1=\tau$ and $\Delta_2=1+\tau$ between adjacent points of the quasicrystal
$\Lambda_{\tau}([0,1))$, this means that all quasicrystals in this case will be applicabel for the classical Vernam cipher.

Using the zero point as the seed point, we obtain the model set
\begin{gather*}
\{\ldots,\!-\!1\!-\!3\tau,
\!-\!2\tau,\!-\!\tau,0,1\!+\!\tau,1\!+\!2\tau,2\!+\!3\tau,
2\!+\!4\tau,3\!+\!5\tau,\ldots\}
\end{gather*}
Rewritten in terms of minimal distances between quasicrystal points it is
\begin{gather*}
\{\ldots,\Delta_2,\,\Delta_1,\,\Delta_1,\,\Delta_2,\,\Delta_1,\,\Delta_2,\,\Delta_1,\,\Delta_2,\ldots\}.
\end{gather*}
The respective one-time pad is
$\{\ldots,1,\,0,\,0,\,1,\,0,\,1,\,0,\,1,\ldots\}.$

It may happen that some point $x'$ is arbitrary close to an end point of $\Omega$.
Then it would be a time demanding computational task to decide whether $x'\in \Omega$ or $x'\notin \Omega$.
Such a difficulty is simply avoided by disqualifying from our consideration any point $x'$
which comes closer to the boundary of $\Omega$ than the distance $\varepsilon$ agreed in advance.
This will slightly modify the algorithm, but we still can define which distance is marked by $0$ and which one is marked~by~$1$.

The real numbers of the form $a+b\tau$ can be understood as given by two integer components $(a,b)$.
Since our operations are only addition or subtraction, the arithmetics of such numbers is elementary introduced.
Consequently all transformations are performed with absolute precision.

In higher dimensions, it is adequate for our purposes to take for the quasicrystals a straightforward concatenation of one-dimensional ones in pairwise orthogonal directions
or consider irrational projections of different higer-dimensional lattices.

\section{Quasicrystal as an one-time pad and its short key}\label{sec_new_Vernam}
The general algorithm of the core of the modified Vernam cipher
\begin{itemize}
  \item represent open data in binary system by a set of a length $n$;
  \item randomly choose input data to create one-dimensional quasicrystal with two tiles only (irrationality and window);
  \item randomly choose a seed point, direction, and order of construction of quasi-crystal points;
  \item construct $n$ points of quasicrystal and write one-time pad as a set of minimal distances coded by zeroes and units;
  \item apply summation modulo two to binary open data and one-time pad;
  \item transmit the encrypted data and a short key or a part of it.
\end{itemize}

Cryptovariables of a key that are randomly chosen:
\begin{itemize}
  \item $m$, $\{1,2\}$ --- fix $\tau$ and $\tau'$ and their order by choice of integer $m$  (\mbox{$x^2=mx+1$}, $m=1,2,\ldots$ or $x^2=mx-1$, $m=3,4,\ldots$);
  \item $a$, $b$ --- integers, fixing a seed point of the quasicrystal;
  \item $c$ and $d$  are arbitrary real numbers, fixing the position and size of the acceptance window $\Omega$;
  \item $\{`+', `-'\}$ --- direction of quasicrystal construction (left or right from the seed point);
  \item $\{1,2,3\}$ sequence of distances of checked adjacent points;
  \item $\varepsilon$ is the maximal distance from the end points of $\Omega$ (to speed up the computations).
\end{itemize}
In this way, we indicated all the parameters on which the quasicrystal depends, they form an infinite set. However, it is possible to use a modified algorithm in which part of the key will be fixed and will not be transmitted with the encrypted message, for example, the coordinates of the starting point or the direction of construction of the quasi-crystal can be fixed. In this case, the size of the part of the transmitted key will be reduced and the encryption will become asymmetric, since the keys used for encoding and the transmitted keys will not match.

\section*{Conclusions}
In the paper, one of the simplest versions of the quasicrystal-based one-time pads is described.
Despite the fact that a one-time pad is not randomly generated, but is built algorithmically
and may resemble a stream cipher, the set of input data for its construction is still random and belongs to a set of infinite cardinality.
Even in this simplest form, it appears adequate for many applications.
It is obvious that the key can be generalized to the case of three different minimum distances,
which corresponds to the summation of integers modulo three in the ring $\mathbb{Z}_3$.

Our proposed approach avoids shortcomings of the standard Vernam cipher, such as the key storage and transmission problem, which is usually caused by the fact that the key length is equal to the message length.

The next steps in research should be a practical implementation of the Wernam cipher with a key constructed using a quasi-crystal,
an application to data encryption, and a separate rigorous study of the stability of the updated Wernam cipher.

More elaborate encoding algorithms can be built by exploiting
properties of higher dimensional quasicrystals.
The modified Vernam cipher can also be applied to encoding not only text but other data, such as raster images.
There are also other applications of the quasicrystals e.g. for generation of
random numbers~\cite{Guimond-Patera-Patera2003},
for chaotic permutation of data, which allows to evenly distribute lost information or can be used as a permutation cipher~\cite{Nesterenko2009}.
\subsection*{Acknowledgment}
M.N. is grateful for the hospitality extended to her at
Czech Technical University in Prague during the work on this project.
Work by M.N. was partially supported by a grant from the Simons Foundation (1290607).
\bibliographystyle{spmpsci}

\end{document}